\documentclass{article}

\usepackage{PRIMEarxiv}

\usepackage[utf8]{inputenc} 
\usepackage[T1]{fontenc}    
\usepackage{hyperref}       
\usepackage{url}            
\usepackage{booktabs}       
\usepackage{amsfonts}       
\usepackage{nicefrac}       
\usepackage{microtype}      
\usepackage{lipsum}
\usepackage{fancyhdr}       
\usepackage{graphicx}       
\graphicspath{{media/}}     

\newcommand{\beginsupplement}{%
    \setcounter{table}{0}
    \renewcommand{\thetable}{S\arabic{table}}%
    \setcounter{figure}{0}
    \renewcommand{\thefigure}{S\arabic{figure}}%
 }

\title{Digital staining in optical microscopy using deep learning - a review
\thanks{\textit{\underline{Citation}}: 
\textbf{Authors. Title. Pages.... DOI:000000/11111.}} 
}

\begin{document}

\begin{flushleft}
{\Large
\textbf\newline{Digital staining in optical microscopy using deep learning - a review}
}
\newline
\\
Lucas Kreiss\textsuperscript{1,2*}, 
Shaowei Jiang\textsuperscript{3}, 
Xiang Li\textsuperscript{4}, 
Shiqi Xu\textsuperscript{1}, 
Kevin C. Zhou\textsuperscript{1,5}, 
Alexander Mühlberg\textsuperscript{2}, 
Kyung Chul Lee\textsuperscript{1,6}, 
Kanghyun Kim\textsuperscript{1}, 
Amey Chaware\textsuperscript{1}, 
Michael Ando\textsuperscript{7}, 
Laura Barisoni\textsuperscript{8}, 
Seung Ah Lee\textsuperscript{6}, 
Guoan Zheng\textsuperscript{3}, 
Kyle Lafata\textsuperscript{4},
Oliver Friedrich\textsuperscript{2} and 
Roarke Horstmeyer\textsuperscript{1}
\\
\bigskip
\textsuperscript{1}Department of Biomedical Engineering, Duke University, Durham, NC 27708, USA\\
\textsuperscript{2}Institute of Medical Biotechnology, Friedrich-Alexander University (FAU), Erlangen, Germany\\
\textsuperscript{3}Department of Biomedical Engineering, University of Connecticut, USA\\
\textsuperscript{4}Department of Radiation Physics , Duke University, Durham, NC 27708, USA\\
\textsuperscript{5}Department of Electrical Engineering \& Computer Sciences, University of California, Berkeley, CA, USA\\
\textsuperscript{6}School of Electrical \& Electronic Engineering, Yonsei University, Seoul, 03722, Republic of Korea\\
\textsuperscript{7} - add -\\
\textsuperscript{8} Department of Pathology, Duke University, Durham, NC 27708, USA \\
\bigskip
\textsuperscript{*} \href{mailto:lucas.kreiss@duke.edu}{lucas.kreiss@duke.edu} or \href{mailto:lucas.kreiss@fau.de}{lucas.kreiss@fau.de}

\end{flushleft}

\textbf{Abstract:} 
Until recently, conventional biochemical staining had the undisputed status as well-established benchmark for most biomedical problems related to clinical diagnostics, fundamental research and biotechnology. Despite this role as gold-standard, staining protocols face several challenges, such as a need for extensive, manual processing of samples, substantial time delays, altered tissue homeostasis, limited choice of contrast agents for a given sample, 2D imaging instead of 3D tomography and many more. Label-free optical technologies, on the other hand, do not rely on exogenous and artificial markers, by exploiting intrinsic optical contrast mechanisms, where the specificity is typically less obvious to the human observer. Over the past few years, digital staining has emerged as a promising concept to use modern deep learning for the translation from optical contrast to established biochemical contrast of actual stainings. In this review article, we provide an in-depth analysis of the current state-of-the-art in this field, suggest methods of good practice, identify pitfalls and challenges and postulate promising advances towards potential future implementations and applications.

\textbf{Keywords: Deep learning, digital staining, optical microscopy, virtual staining, in-silica, pseudo-H\&E, virtual fluorescence, generative models}

\section{Introduction}
\label{sec:intro}
Biomedical sciences heavily rely on numerous biochemical staining protocols to achieve specific cell identification or tissue classification. Chemical binding between target molecules and engineered molecular markers, loaded with artificial contrast agents, can create artificial yet specific contrast for optical microscopy. Due to their molecular specificity, these staining protocols are the established benchmark for most biomedical problems related to clinical diagnostics, fundamental research, and biotechnology. Beyond their undisputed success, these labeling-intense protocols still require extensive processing of the samples, which can cause substantial time delays, affect tissue homeostasis, might limit the choice of available contrast agents and often only allow 2D imaging of tissue slices instead of 3D tomography. The most common histological stain of hematoxylin and eosin (H\&E), for instance, is based on tissue embedding, fixation (Formalin-Fixed Paraffin-Embedded - FFPE), manually slicing into thin sections (typically 3-7\textmu{m}) and staining, before imaging under conventional microscopes or whole slide scanners can be pursued. 
\newline
\begin{figure}[ht!]
\begin{center}
\includegraphics[width=\linewidth]{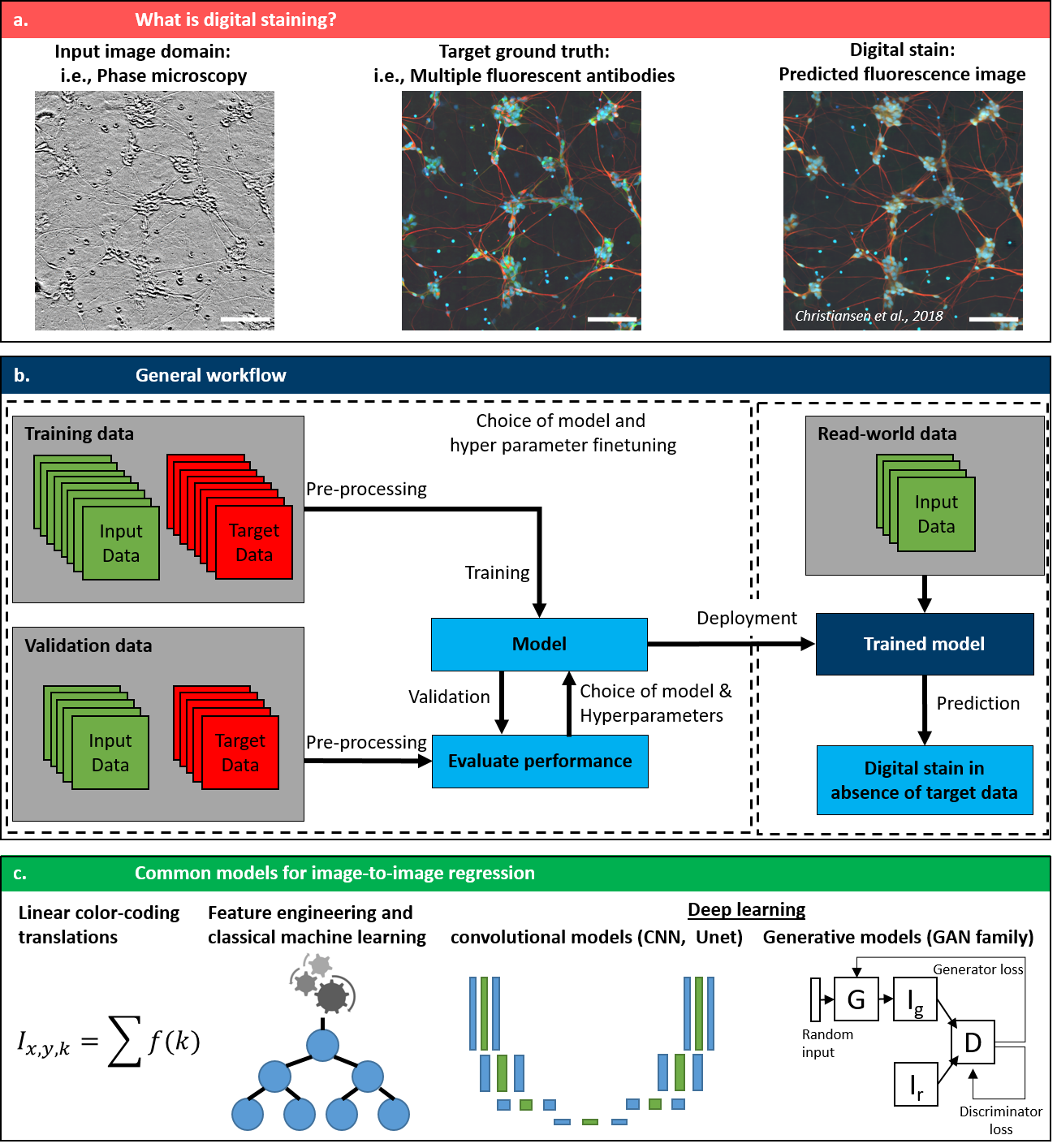}
\end{center}
\caption{\textbf{Basic principle of Digital staining} (a) Example data showing phase contrast microscopy images as input to predict fluorescence images from specific antibody staining as ground truth. Images show human motor neurons stained for nuclei (DAPI), dendrites (anti-MAP2) and axons (anti-neurofilament). Ground truth image were acquired on confocal fluorescence microscopy (scale bar: 100~\textmu m). Data available at \url{https://github.com/google/in-silico-labeling} from Ref.~\cite{60} with permission from Elsevier and Copyright Clearance Center. (b) The general supervised machine learning workflow for digital staining. (c) The most commonly used models: besides earlier implementations of color-coding with a linear contrast translation equation f(k) or feature engineering and classical ML, almost all modern digital staining implementations use deep learning with either CNN and GAN architectures.}
\label{fig:fig1}
\end{figure}

Label-free optical technologies, on the other hand, exploit natural contrast mechanisms, instead of relying on a limited choice of exogenous markers in the above mentioned staining procedures. Simple white-light microscopy, for instance, rely on amplitude differences based on scattering and absorption properties of cells and tissues, optical phase microscopy measures phase contrast based on refractive index (RI) differences, birefringence, or orientation, while other imaging modalities use intensity or lifetime of natural autofluorescence (AF). Although these label-free contrast mechanisms can actually carry highly-relevant information related to factors like density and thickness, mass, redox-ratio and many more, their specificity as direct biomarkers is typically less obvious to the human observer. 
Over the past decades, machine learning (ML) or artificial intelligence (AI) demonstrated vast success in optical microscopy, e.g. in automated detection of diseases~\cite{aresta2019_bach}, 3D image segmentation~\cite{haft2019_deep} or simultaneous optimization of microscopy and software components~\cite{muthumbi2019_jointly_optimized}. In conventional pathology, AI models are often used to perform classification or segmentation of histology images from diseased and healthy tissues. As common in most supervised ML, training of these models requires large datasets with reliable ground truth labels. These labels are commonly generated manually by experts, i.e., for the automated segmentation of background, cell boundaries, and cell compartments by convolutional neural networks (CNN)~\cite{chen2014_cell_segmentation}. Especially with the rise of the U-Net architecture~\cite{ronneberger2015_Unet} for image segmentation, cell segmentation could be solved more effectively. Nevertheless, the conventional procedures of histological staining and manual annotations are still rather time-consuming, and the need for reliable ground-truth data often acts as bottlenecks for throughput in digital pathology.
\newline
Over the past decade, ML researchers have developed several techniques for image-to-image translation. Upon training and validation, these generative models allowed transfer from one image domain to another, e.g., from maps to satellite images~\cite{liu2017_unsupervised_map2satelite}, from horses to zebras~\cite{zhu2017unpairedcylcleGAN} or for style transfer in art~\cite{DomainTransfer_Art}. Recently, alternative image-to-image training strategies, such as Normalizing Flows~\cite{kingma2018glow,kobyzev2020normalizing} and Denoising Diffusion Probabilistic Models~\cite{ho2020denoising,saharia2022palette} have also gained significant popularity. Digital staining (DS) is an emerging concept in the field of computational microscopy that can digitally augment microscopy images by transferring the contrast of input images into a target domain. Implementation of digital models is most often based on machine learning algorithms, that are trained on pairs of input and target images. In a nutshell, these ML models then learn to link characteristic features in structure and contrast from one input domain (most often a label-free image) with those of the target domain (most often images of well know molecular specificity). Thereby, digital staining very elegantly bypasses two obstacles. During the development and training of computational models, digital staining omits the need of manual annotations of ground truth data, by obtaining the ground truth annotations from specific stainings. Upon deployment, the inference with a trained model can then circumvent the time-consuming and tedious procedure of actual sample preparation, including sectioning and staining. 
\newline
Despite the vast potential of this technique, the growing number of new digital staining pipelines and a wider range of applications, thorough review articles on this topic are rare and only touch side aspects of digital staining. A 2022 review on GANs in ophthalmology~\cite{GANs_ophthalmology} mentioned some DS techniques in the specific use case of transforming fundus photographs to angiography images. 
Jiang et al. provide a concise review on deep learning (DL) in cytology, including classification, segmentation, object detection and stain normalization of microscopy images~\cite{Computational_Cytology_review}, but without covering digital staining as such. In a similar fashion, Wu et al., touch on the topic of style transfer in microscopy in their 2021 review on computational histopathology~\cite{Computational_Histopath}, however only in the context of color normalization. In 2018, Jo et al., mentioned 'image enhancement via style transfer' as promising developments for the specific technique of quantitative phase imaging (QPI)~\cite{jo2018_QPI_and_AI_review}, but without generally reviewing the entire field of digital staining. Rivenson et al., published a 2020 review article on virtual staining for histopathology~\cite{Rivenson2020_Review}. However, it only targeted digital staining of FFPE sections and did not include the immense increase in publications in this field over the past three to four years (see supplementary fig.~\ref{fig:sup_fig3}~A). Latest reviews from 2022 and 2023 summarized the concept to translate input images into target images in the sole context of histological tissue section~\cite{pillar2022_shortReview_AF,bai2023deep}, but without an in-depth analysis of other digital staining applications, including (live) cell staining.

\section{Basic principle and key examples}
\label{sec_basic_principle}
Successful implementation of digital staining essentially relies on four key parts:
\begin{itemize}
    \item the use of \textbf{input images} that carry a sufficient amount of information to allow the translation into the target domain (seechapter~\ref{section:input_contrasts}). This input domain usually relates to the use of a label-free technique, but it is not limited to that.
    \item the use of \textbf{target images} that show a reliable ground-truth information that can be linked to the features in the input domain (seechapter~\ref{sec_target_domain}). These target images usually use the biochemical specificity of molecular stains as ground-truth, but are not restricted to those.
    \item the use of appropriate \textbf{computational models} that can translate input images to target images (see chapter~\ref{sec_computational_models}). Most often, this image-to-image regression problem is solved by machine learning algorithms (specifically U-Net or GAN architectures), but earlier implementations also relied on linear, mathematical formulas to translate color spaces.
    \item a procedure to accurately generate \textbf{paired input and target images}  (see chapter~\ref{section:generate_pairs}). The exact registration of input pixels and target pixels of the same structures might seem trivial but is essential to enable the model to perform accurate image-to-image regression. While a few recent implementations use unsupervised learning with unpaired images for training, all implementations at least require input and target image pairs for a truthful validation of the predictions, as discussed below.
\end{itemize}

Thus, digital staining can be viewed as a holistic concurrence of biology, optical microscopy and computational modeling. Successful implementations rely on an understanding of the entire workflow that starts from a reasonably posed biological problem, involve input images that carry a sufficient amount of information with respect to the biological task, as well as target images that can be linked to the information from the input domain and end with a computational model that is able to accurately translate input images into target images. Furthermore, the practical workflow to generate pairs of input and target images as well as the choice of quantitative metrics for training and validation are important considerations for digital staining.
\newline
Depending on the mode of operation and the preference of the authors, the concept of DS is also termed 'virtual staining', '\textit{in silico} staining', 'pseudo-H\&E staining' or 'virtual fluorescence'. 

The earliest, and still one of the most common, implementation of DS translates label-free images of tissue section into target images of well-known and widely accepted histological stainings. This is often based on two subsequent tissue sections, where one is imaged with label-free modalities as input, while the consecutive section is used for a conventional histology stain as target. This digital H\&E staining has been shown extensively and for a multitude of different organ samples based on label-free autofluorescence~\cite{29}. 
\newline
In 2018, Christiansen et al. demonstrated the next stage for digital staining from live cell cultures with different IF dyes in a shared optical path, by using phase microscopy and a U-Net model~\cite{60}. The use of fluorescently labelled antibodies for digital staining in live cells opened the door for many new biomedical experiments, like an extension into 3D digital staining~\cite{58}, the use of digital staining to promote prior-informed cell segmentation~\cite{70}, digital staining of two different cell cycle markers for mitosis stage classification~\cite{74} or a detailed evaluation of virtual labeling of mitochondria in living cells~\cite{118,135}.
\newline
Besides the conceptual advancements of digital staining, the field was undoubtedly fueled by the introduction of more powerful ML models for image-to-image regression, such as U-Net~\cite{ronneberger2015_Unet}, generative adversarial networks (GAN)~\cite{goodfellow2014_GANs} and cycle conditional GANs~\cite{zhu2017unpairedcylcleGAN}. Since these models became generally more available and were applied to digital staining, e.g., the use of the 'Pix2Pix' for digital staining in 2018~\cite{20} or the stainGAN, which was initially used for stain normalization~\cite{shaban2019staingan}, the number of publications in this field increased exponentially over the past 3-4 years (see supplementary fig.~\ref{fig:sup_fig3}~A). 

\section{Input domains: label-free contrast mechanisms in optical microscopy as "optical specificity"} 
\label{section:input_contrasts}
As mentioned above, label-free contrast mechanisms often carry highly-relevant information that can be linked to functional and/or morphological features density, thickness or mass, the redox-ratio of a cell cycle, surface topography, presence or absense of certain molecules and many more. Wheather this information is sufficient for a given digital staining task, is among the first and most important considerations when implementing a digital staining model, as discussed in section~\ref{Diss_goodpractice} below.
\newline
In this chapter, all reviewed publications are categorized according to contrast mechanism of the input domain. Label-free microscopy techniques are commonly used to generate input images, while elaborate staining procedures of known biochemical specificity are usually used as target images. The two label-free techniques of optical phase contrast and wide-field / white light illumination are the most commonly used techniques to generate input images with 18.6\% and 16.8\% of our reviewed literature respectively. Other notable label-free input imaging methods include autofluorescence (AF), nonlinear techniques, or photoacoustic imaging. There are several studies that employ combinations of different contrasts. On the one hand, this was implemented \textit{in one single setup}, e.g., in Fourier pythography microscopy (FPM) as a combination of amplitude and phase contrast~\cite{41}, in dark field reflectance and autofluorescence (DRUM)~\cite{131} or in complementary nonlinear techniques~\cite{18,50,66}. On the other hand, some papers present the use of \textit{different imaging systems} for combined data input, e.g., wide-field and phase contrast~\cite{58,60,106}. Furthermore, there are also several implementations of stain-to-stain translation, where inputs from one stain are digitally transferred to a different target stain.

\begin{figure}[ht!]
\begin{center}
\includegraphics[width=\linewidth]{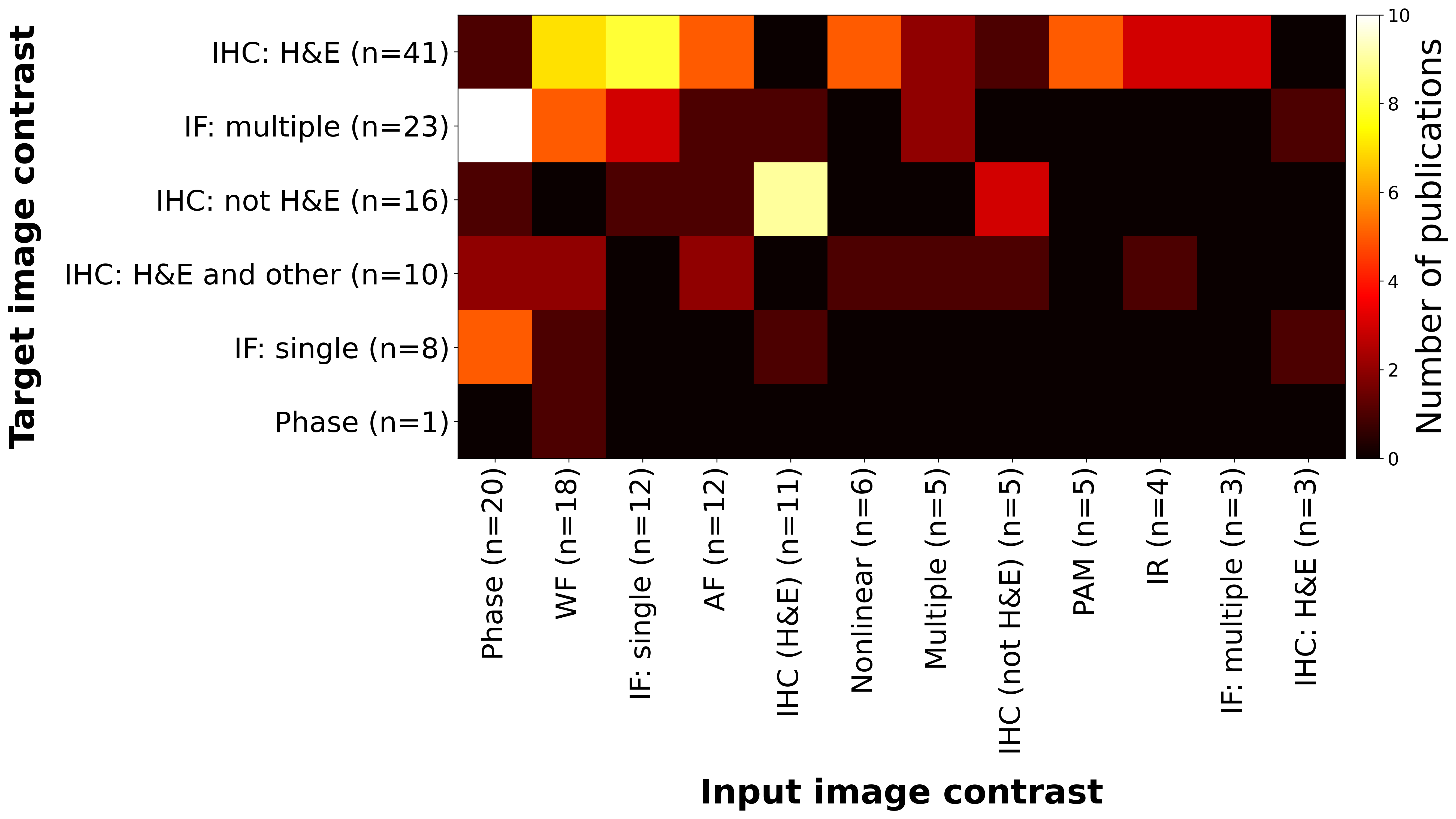}
\end{center}
\caption{\textbf{Pairings of input and target contrast} Target image contrast is plotted against the input contrast, with the number of publications for each combination as a color-coded heat map. IHC = immuno-histochemcial stain, IF = immuno-fluorescence stain, WF = wide field (white light illumination), AF = autofluorescence, PAM = photo-acoustic microscopy, IR = infra-red. An extended version including a detailed literature analysis can be found in the supplementary material of this manuscript, see suppl.fig.~\ref{fig:sup_fig2}}
\label{fig:fig2}
\end{figure}

While H\&E is the most wide-spread stain used for digital staining, the single most popular combination is the use of phase contrast microscopy as input images to predict multiple IF stains, as displayed in Figure~\ref{fig:fig2}. Since most of these IF stains are targeting membrane (Dil stain), nuclei (DAPI or Hoechst) or cytoskeleton (Microtubuli, MAP-stains), phase imaging techniques are an ideal match, as their optical phase contrast is highest for the very same cellular structures (membranes, nuclei and cytoskeleton). 

The most important label-free optical techniques are briefly presented in this chapter, while biochemical staining methods which are usually used as target images, are presented in the following chapter~\ref{sec_target_domain}. 

\subsection{Wide-field (WF) microscopy}
\label{section:input_contrast_WF}
Perhaps the most basic type of optical microscope is the standard wide-field microscope, known since the beginnings of optical microscopy. The basic design consists of a light source that illuminates an extended area typically of a thin sample which scatters and transmits a fraction of the illumination into a lens or collection of lenses that image the light onto an arrayed detector. Wide-field microscopy has also been referred to as bright-field microscopy and white-light microscopy, among others, stemming from the particular illumination conditions, which we will discuss shortly. 

In its most basic form, wide-field microscopy offers qualitative contrast derived from the spatially-varying complex transmittance of the sample: 
\begin{equation}\label{complex_transmittance}
    t(x,y) = \exp(j2\pi n(x,y)\Delta z(x,y)/\lambda)
\end{equation}
where $n(x,y)$ is the spatially-varying complex refractive index of the sample, $\Delta z(x,y)$ is the sample thickness, and $\lambda$ is the wavelength of the illumination.
The real part of the refractive index imparts a phase shift on the incident light that is often difficult to observe in thin samples using standard bright-field illumination, or illumination whose angular range falls within the numerical aperture (NA) of the objective lens. However, off-axis illumination in the bright-field regime or in the dark-field regime (i.e., with illumination angles higher than the cutoff imposed by the NA of the objective) can highlight certain features that may be used for virtual staining, such as cell or organelle boundaries. The imaginary part of the refractive index corresponds to absorption induced by the sample. As such, wide-field microscopy can be useful for imaging certain types of cells that contain strongly absorbing molecules at certain wavelengths, such as red blood cells and melanocytes. The wavelength dependence of the absorption is often useful for distinguishing certain types of molecules, which can be achieved with a wide-field microscope by sweeping the illumination wavelength or by using white-light illumination with a multi- or hyperspectral camera.

For thicker samples, a simple model based on complex transmittance map (Eq.~\ref{complex_transmittance}) is insufficient. Such samples may exhibit higher attenuation contrast due to multiple scattering and absorption, which can be quantified by an attenuation coefficient, $\mu_t$, that subsumes both the absorption coefficient, $\mu_a$, and scattering coefficient, $\mu_s$, though a standard wide-field microscope generally cannot distinguish the effects of the two sources. While such coefficients are gross or bulk metrics of biological samples (i.e., having an opaque relationship with the 3D structure of the sample), they can still offer useful sources of contrast for virtual staining~\cite{58}.


\subsection{Phase sensitive methods}
Phase contrast (PC) is an important endogenous contrast mechanism of label-free samples. Small changes in the refractive index and thickness of cells result in detectable changes in the optical phase. Generally, phase contrast microscopy attenuates the background light and compensates the phase shift of the scattered light. This way, the scattered light interferes with the background light more constructively, which enhances the image contrast~\cite{park2018quantitative}. Phase microscopy techniques are quite diverse in their exact implementation. They range from the use of phase rings or spatial light modulators, to interferometric setups or active illumination control and most often include computational phase reconstruction. 

Phase contrast microscopy and differential interference contrast (DIC) microscopy are still two of the most commonly used phase imaging modalities that reveal structures of semi-transparent cells that are invisible to the previously discussed wide-field microscopy. Due to the substantial development of PC and DIC in the last half-century and the increasing demand for monitoring \textit{in vitro} cells, those two modalities are now commonly available in commercial microscope solutions. Therefore, a large and diverse amount of PC and DIC studies have been conducted on multiple sites for predicting fluorescence labels including nuclei and dendrites for human motor neurons cells, as well as nuclei and membranes for human breast cancer line cells~\cite{60}. Further, DIC-based virtual staining has been proposed in hematology to replace the laborious and inconsistent H\&E stain of blood smears~\cite{52}. In this case, as DIC only preserves the edges of phase images, they tend to lack details for accurate predictions of the inner-cellular structures. To relieve this issue, Tomczak et al.~\cite{52} proposed to add an auxiliary task of nucleus and cytoplasm segmentation in addition to the prime domain transformation task (i.e., to predict H\&E stain from DIC images), which forces the encoder to be aware of the shape of structures. Compared to transformation networks trained with the prime domain transformation task alone, such a multi-task learning method can improve performance on digitally staining leukocytes from hematology slides imaged with DIC.

Another imaging technique based on the RI of the sample is optical coherence tomography (OCT)~\cite{zhou2021unified}. Modern point-scan OCT is typically implemented in the frequency domain with a Michelson or Mach-Zehnder interferometer, using wavelength-swept light sources or broadband (low-coherence) sources such as superluminescent diodes for illumination. In analogy to ultrasound imaging, OCT uses an optical ``pulse-echo'' time-of-flight method to create tomographic line-scan images along an optical ray, which can penetrate up to a few millimeters inside human tissue. Scanning mirrors can then be used to move the optical beam to transversely across the sample and create a volumetric 3D image of a tissue sample. While the lateral resolution of OCT depends on the numerical aperture (NA), its axial resolution is inversely proportional to the bandwidth of the source~\cite{drexler2015optical}. Since its invention in the early 1990s, OCT has become one of the most successful optical methods in the medical industry~\cite{drexler2015optical}. Due to this commercial success, OCT devices are now available off-the-shelf. For instance, Lin et al. use a multi-modal OCT system (AcuSolutions Inc, Taiwan) that can create registered images from both optical coherence microscopy and fluorescence microscopy~\cite{65}. The images from the two modalities were merged and false-colored to create pseudo-H\&E images. Extensive in-depth comparisons between pseudo-H\&E image and frozen-section H\&E image from various biopsy specimens are provided to show that the proposed digital stain method can provide H\&E images that describe cellular-level morphology around two times faster than the frozen-section method~\cite{65}. In addition, another study shows that digital staining can also be achieved from \textit{in vivo} OCT measurements~\cite{16} where tomographic images of the optic nerve heads are acquired from 10 healthy subjects using a standard OCT eye scanner (Heidelberg Engineering Inc, Germany). Four different tissue types are identified based on pixel-intensity histograms and digitally stained in a way that connective and neural tissues of the optics nerve heads can be easily visualized~\cite{16}. 

While the well-established techniques of phase-contrast microscopy and DIC provide qualitative phase contrast by converting phase differences into intensity differences, quantitative phase imaging (QPI) can provide intrinsic quantification of the optical path lengths difference which is a function of refractive index (RI) and sample thickness~\cite{wang2011_SLIM}. Thus QPI shows decent specificity in the imaging signal without requiring any sample preparations. Due to its ability to map the physical refractive index of the sample, digital staining based on QPI has been widely explored recently with various computational microscopy implementations~\cite{park2018quantitative}. The QPI concept was gradually extended towards 3D imaging, which resulted in the invention of gradient light interference microscopy (GLIM) in 2017~\cite{nguyen2017gradient}. GLIM uses data post-processing for filtering of out-of-focus components for 3D imaging. In 2020, this technique was further augmented by computational specificity (phase imaging with computational specificity - PICS) to digitally stain 3D GLIM images using a U-Net implementation~\cite{33}.

Fourier Ptychographic Microscopy (FPM) is a computational microscopy technique that enables wide-field and high-resolution QPI without any interferometry and mechanical scanning~\cite{konda2020fourier}. Usually, a low-magnification lens is used for a wide field-of-view, and an LED array is utilized for varying illumination angles. In FPM, multiple measurements are captured by varying illumination angles, and each measurement represents a different spatial frequency of the sample. Phase information is then recovered via phase retrieval algorithms, that utilize overlapped spatial frequency as a constraint. FPM was already used for digital staining of antibody conjugates stained mouse kidney slides from monochromatic phase images reconstructed with Fourier Ptychography~\cite{41}. An FPM-like setup using the same active illumination of a LED array was also used to digitally stain cell membrane, and nuclei in two different cell cultures~\cite{42}, although actual FP reconstruction was not applied in that case.

\subsection{Autofluorescence (AF)}
There are several naturally occurring proteins, that emit fluorescence upon excitation by UV or blue light. This process of autofluorescence is often exploited for label-free fluorescence imaging. The most common autofluorescent molecules are listed in table~\ref{tab:autofluorescence}. The excited molecule can then emit standard fluorescence after internal energy conversion (Stokes shift). Intensities, as well as ratios of different autofluorescence molecules can be specific to certain cell types and/or functional states~\cite{croce2014autofluorescence}. Thus, AF is a reasonable candidate to be used for digital staining. Similar to WSI with white light illumination, some articles use whole slide scanners with UV light to excite these natural fluorophores to WSI based on AF contrast~\cite{73}. 

\begin{table}[ht!]
    \centering
    \caption{Most common fluorophores for natural autofluorescence, according to~\cite{croce2014autofluorescence}.}
    \begin{tabular}{l|r|r}
         Fluorophore  & Excitation & Emission \\
                    & wavelength (nm) & wavelength (nm) \\
         \hline
         Phe, Tyr, Trp & 240 - 280 & 280 - 350 \\
         Cytokeratins & 280 - 325 & 495 - 525 \\ Collagen & 330 - 340 & 400 - 410 \\
         Elastin & 350 - 420 & 420 - 510 \\
         NAD(P)H & 330 - 380 & 440 - 462 \\
         Flavins & 350 - 370, 440 - 450 & 480, 540 \\
         Fatty acids & 330 - 350 & 470 - 480 \\
         Vitamin A & 370 - 380 & 490 - 510 \\
         Porphyrins & 405 & 630 - 700 \\
         Lipofuscins & 400 - 500 & 480 - 700 \\
    \end{tabular}
    \label{tab:autofluorescence}
\end{table}

\subsection{Nonlinear Techniques}
Optical, nonlinear label-free contrast mechanisms described here include multiphoton microscopy (based on nonlinear AF and second harmonic generation - SHG) and Coherent Anti-Stokes Raman Scattering (CARS).

Multiphoton Microscopy can be used for label-free imaging using natural AF and other nonlinear  discussed below. Although the non-linear excitation process is slightly different to the single-photon AF, described above, most molecules displayed in table~\ref{tab:autofluorescence} can also be excited with a corresponding two- or three-photon excitation. Compared to conventional fluorescence, which uses blue or UV light of around 400~nm, MPM uses longer wavelengths typically in the range of 780-850~nm (two photon process) or 1,100-1,300~nm (three photon process). This avoids the strong scattering and absorption of biological tissues in the UV range and is not yet affected by the immense attenuation from absorption in water towards the far infra-red region. Therefore, MPM enables deeper tissue imaging than single-photon microscopy. Additionally, the signal generation is naturally limited to the confined focal volume, which outwears the need for a pinhole in the detection path. Most commonly, the native fluorophores of NADH and flavins, are used for label-free MPM~\cite{Zipfel2003}. Similar to single-photon autofluorescence, this signal was shown to be specific for certain cell types and/or functional states~\cite{lemire2022natural,gehlsen2015non}, making it a useful input contrast for digital staining. 
\newline
Furthermore, MPM naturally enables higher harmonic generation (second harmonic generation - SHG or third harmonic generation - THG) as additional contrast mechanism for imaging. SHG or THG are based on the electrical field component of the incident light and the polarization properties of the sample. This electrical field induces a directional polarization within the sample, which in turn induces the emission of a secondary wave at higher frequency. In contrast to fluorescence, SHG or THG does not experience a Stokes shift. This signal is very specific to structures within the sample that have respective non-linear susceptibility properties (i.e., $\chi^{(2)} > 0$ for SHG or $\chi^{(3)} > 0$ for THG). SHG for instance, is specific for structures that lack inversion symmetry ($\chi^{(2)} > 0$), such as biological molecules of collagen, myosin and tubulin~\cite{Schurmann2007}. 
\newline
A multi-modal microscopy system, including coherent anti-stokes Raman scattering (CARS) at 2,850~cm$^{-1}$, SHG in forward direction and two-photon AF in backward direction~\cite{Heuke2013}, was used to demonstrate a computational transformation from images label-free multi-modal contrast to image with an artificial H\&E contrast~\cite{66}. This translation was later updated by using GAN models~\cite{50}.

\subsection{Photoacoustic microscopy (PAM)}
Photoacoustic imaging is based on the photoacoustic effect~\cite{Rosencwaig1980} and detects sound propagation upon laser excitation of the most prominent absorbers in tissue~\cite{Xu2006}. Thus, PAM promises high molecular specificity to molecules that have a high absorption coefficient, such as hemoglobin, water, melanin and collagen~\cite{Xu2006}. As ultrasonic scattering is typically weaker in tissue compared to optical scattering, photoacoustic microscopy can produce absorption images at deeper depths compared with traditional microscopy techniques, which makes it suitable for a variety of \textit{in vivo} studies~\cite{yao2013photoacoustic}.
 Digital staining of PAM images was demonstrated for FFPE brain secitons~\cite{55,116} or skin sections~\cite{115,119}.

\section{Target domain: biochemical stains as ground-truth}
\label{sec_target_domain}

While the previous chapter~\ref{section:input_contrasts} discusses label-free optical imaging techniques, that are mostly used as input images for digital staining, this chapter presents a similar analysis for artificial staining methods that are usually used as target images for digital staining. Here, we have grouped the typical staining methods into: immuno-histochemical staining (IHC) and immuno-fluorescence staining (IF).

\subsection{Histological stains}
IHC or histological staining methods first rely on the extraction of biopsy samples. There are several different techniques for taking of biopsies, like strip biopsy~\cite{Karita1991}, by using endoscopic pincer grasping instruments~\cite{Stefanchik2010} or ligating devices~\cite{Akiyama1997}. These samples are then embedded and sectioned. There are two main approaches for tissue embedding: embedding in paraffin~\cite{Wright1990} or snap freezing in optimal cooling temperature gel~\cite{Mager2007}. Each of these procedures comes with certain procedural requirements and different time duration. 
\par
Depending on the type of embedding, the samples are then sectioned by cryotomes or microtomes to thin slices, typically between 3 and 10~\textmu m. Finally, these sections are mounted on glass slides and stained. In the case of H\&E staining, following Cardiff et al.~\cite{Cardiff2014}, paraffin tissue sections are first cleared of paraffin in baths of xylene (three changes for 2 min per change), then hydrated by ethanol baths (three changes of 100\% ethanol for 2 min per change, transfer to 95\% ethanol for 2 min, transfer to 70\% ethanol for 2 min) and rinsed in running tap water (2 min)~\cite{Cardiff2014}. 
\newline
Afterwards, the tissue sections are stained in hematoxylin solution (3 min), washed again in running tap water (5 min) and then stained with eosin (2 min)~\cite{Cardiff2014}. The samples are  dehydrated (dipping in 95\% ethanol, transfer to 95\% ethanol for 2 min, two transfers to 100\% ethanol for 2 min per change) and cleared in three changes of xylene (2 min per change)~\cite{Cardiff2014}. Thereby, hematoxylin stains cell nuclei and eosin extracellular matrix and cytoplasm. Finally, the stained tissue sections are sealed and preserved between glass slice and a coverslip~\cite{Cardiff2014}. Thus, the staining protocol alone already accounts for at least 90~min, and the entire procedure from biopsy acquisition to microscopic images of the stained tissue sections can easily last multiple days or even weeks, when considering queuing times in the common laboratory work-flow. In the current state-of-the-art of digital staining, H\&E staining is by far the most common target stain for digital staining, as displayed in figure~\ref{fig:fig2}. 
\newline
Besides this H\&E staining, there is an innumerable amount of other IHC staining procedures, including picro Sirius red staining for collagen~\cite{Whittaker1994}.
Other IHC stains that were already used for digital staining include human cancer marker (Ki-67 antigen)~\cite{18}, Jones’ stain~\cite{28,29,71,112},  Masson’s trichrome~\cite{28,29,71,84,106,107,108,110,112}, picro sirius red~\cite{35,57,93}, orcein~\cite{35},Verhoeff van Gieson (EVG) stains~\cite{57,93}, periodic acid-Schiff (PAS) stain~\cite{102,106,108}. 


\subsection{Immuno-fluorescence staining (IF)} 
The third category is the use of fluorescence markers for staining. This can either be achieved by a fluorescent primary antibody (DAPI)~\cite{Otto1990} or by using the established combination of primary antibodies against specific epitopes and a fluorescent secondary antibody. In the latter case, the primary antibodies are sometimes similar to those used in IHC. IF staining can be used on FFPE sections or frozen sections. Moreover, IF can be used with cell cultures, that are not feasible with tissue staining of IHC stainings. This enables a series of new applications.
\newline
In combination with a shared optical system to generate paired input and target images (see chapter\ref{section:generate_pairs}), IF is the best viable option to perform digital staining for living cells in culture. The most common examples for IF techniques in digital staining include stains for cell membranes~\cite{33, 42, 58,60,91}, DAPI or Hoechst stains for nucleii~\cite{33,42,46,58,60,61,62,69,72,73,91,95,103,104}, Rhodamine B isothiocyanat for viruses~\cite{44}, axons markers (tau stain~\cite{46}, anti-neurofilament stain~\cite{60}, antiMAP2  for dendrites~\cite{46,60,73}, live and dead cell markers (NucBlue as “live” reagent and NucGreen as “dead” reagent or PI as dead cell marker)~\cite{54,60,61,72,73}, actin markers~\cite{56,58,58,61,69,95}, Mitochondria (MitoTracker Red)~\cite{59}, antiTuj1 for neurons~\cite{60}, endosome~\cite{61}, goldi apparatus~\cite{61}, proliferation~\cite{61}, Myelin marker in brain~\cite{69}, markers for the G1 and S stage of the cell cycle~\cite{74}. 

In addition to these direct molecular markers, fluorescence stains can also be encoded by genetic modification of the target organism to achieve expression of fluorescence markers in target components, e.g., in mitochondria~\cite{118,135}.
Furthermore, IF stains are also being used in multiplexed fashion (see Fig.~\ref{fig:fig4}~E) for multiplexed immunofluorescence (mpIF)~\cite{126,127}.

\subsection{Biochemical specificity of target stains}
\label{Targets_specificity}
Although most stains mentioned above are commonly used as 'gold-standard', they are actually not standardized with respect to their biochemcial binding specificity. In the case of IHC stains, the appearance of stained samples severely dependents on the type of stain solution, the exact staining protocol and the quality or age of dyes. This is especially the case for IHC, but also applies to certain IF stains, like the common fluorescent DNA-stain DAPI. In these cases, a standardized specificity value (commonly stated in \%) is not available. 
\newline
For IF stains on the other hand, 
antibody manufacturers occasionally state reference measurements for specificity. However, it is still challenging to standardize the actual biochemical specificity values across different studies, as it is severely affected by the precise biochemical conditions of the experiment and the environment, including pH value, different behavior in medium vs in cells, ligand buffer interaction, temperature or competing binding partners, to name a few. As displayed in table~\ref{tab:biochemcial_specificity}, the stated specificity values can range between 66\% and almost 100~\% for different target molecules. Moreover, this binding specificity even varies for the same molecule, for instance different antibodies target different binding sites (see the example of anti-tau antibodies in table~\ref{tab:biochemcial_specificity}).

\begin{table}[ht!]
    \centering
    \caption{Examples of primary antibodies for immuno-fluorescence staining with reported binding specificity and features examples for digital staining (DS). MAP2 = Microtubule Associated Protein 2, HEK = human embryonic kidney, isoform specificity = "no detectable non-specific binding".}
    \begin{tabular}{l|l|l|r}
          Anti- & Target molecule & Reported binding & Featured DS \\
                 body &                              &              specificity & references \\
                   
         \hline
         
          Anti- & cytoskeleton & isoform specificity & \cite{46,56,58} \\
                $\beta$-actin &  (actin proteins) &  in knockout cells~\cite{tondeleir2012cells} & \cite{61,69} \\
                  & & & \\
          Anti- & dendrites (microtubule ass- & 90\% in Mouse  & \cite{46,60,72} \\        
          MAP2 & embly during neurogenesis) & brain cells~\cite{ABCAM_Anti_MAP2} & \cite{73,91,133} \\
           &         &                          &      \\
        
        Anti- & axons (neuronal microtubule- & 80\% for pT181 binding site & \cite{46,133} \\
        Tau &  associated protein) & isoform specificity for most & \\
                  & &  other sites in HEK cells~\cite{li2020high}                                & \\
                  & & & \\

         Anti-     & cell cycle marker, &          66\% in human blood cells~\cite{antibodies_Ki67}         & ~\cite{61,80} \\
         Ki-67 & proliferation marker, & & \cite{113,121,126} \\
         & marker for neoplastic tissues & & \\
    \end{tabular}
    \label{tab:biochemcial_specificity}
\end{table}

For most histological applications, this is completely acceptable, as long as the stain quality enables pathologists to count cells, determine diseased tissue and make a diagnosis. In the case of IF stains, careful calibration measurements can still enable quantitative analysis from the same laboratory under the same conditions. In the case of digital staining however, this biochemical specificity represents the fundamental uncertainty that defines the upper limit of trustworthiness of a digital staining model. As discussed here, it is thus ill-advised to consider target images of IHC or IF stains as actual ground-\textbf{truth} in the literal sense.

\section{Computational models to transfer input images to target domain}
\label{sec_computational_models}
As already mentioned, the development of image-to-image regression models, like U-Net~\cite{ronneberger2015_Unet}, GANs~\cite{goodfellow2014_GANs} or cycle conditional GANs~\cite{zhu2017unpairedcylcleGAN} fueled the field of digital staining over the past years. Together, these models make up more than 60\% of all reviewed articles here. A short overview of the basic principle of these models is displayed in Figure~\ref{fig:fig1} and elaborated upon below.

\subsection{Pre-processing}
\label{sec_computational_models_preproc}
Before image data can be used to train a digital staining model, several pre-processing steps are often required. Unless a common optical path is used (see chapter~\ref{section:generate_pairs}), digital staining usually requires image registration to ensure optimal pixel overlay between input and target images. As discussed in chapter~\ref{Diss_pitfalls}, this can lead to several challenges, as for instance with sectioning artifacts when using consecutive sections to generate paired images. The generation of image patches is an additional pre-processing step that is very common. Especially, when images are acquired from large-FOV whole-slide imaging (WSI) systems (see section~\ref{section:input_contrast_WF}), slide images are usually cropped into 2,000-20,000 image patches of 256x256~pix² or 512x512~pix² each before training a digital staining model. 

\subsection{Linear color-coding methods for stain transformation}
Training of data-driven machine learning models is the current method of choice as computational model to transfer style and color from input into target images. However, especially earlier studies also used simpler mathematical equations for color transfer that worked reasonably well, but were often not verified quantitatively on a separate validation data set~\cite{7, 10, 9, 11, 84, 14, 85, 23, 24, 75, 65}. Most of them follow a simple color-coding method, i.e., a linear mathematical model based on Gareau et al.~\cite{7}, which was also applied to the previously mentioned OCT images~\cite{65}. Although most of these linear color coding methods were applied for earlier implementations of DS, they were still used as recently as 2022~\cite{131}.

\subsection{Feature engineering and classical machine learning}
In the next phase of digital staining models, researchers quantified engineered image features and exploited them in classical machine learning models. For instance, k nearest neighbor ~\cite{3}, spectral Angle Map (SAM), Nearest neighbor (NN), nearest mean classifier (NearMean)~\cite{5}, random forest~\cite{6,18} or partial least squares regression (PLS)~\cite{66} were used for digital staining problems. Although this approach requires more human-supervised feature extraction and prior knowledge, it can perform very robustly and often generalizes well across different data sets from the same sample under different imaging systems. On the other hand, it is often challenging to transfer it to other samples and can be more labor-intense than deep learning methods.

\subsection{Deep learning}
Deep neural networks (DNNs) refer to neural networks with multiple layers, allowing for the extraction of increasingly abstract features from input data. While early models in the 1940s and 1950s were limited in their ability to learn from data and to scale to larger and more complex problems, the development of backpropagation in the 1980s sparked renewed interest in DL. However, computational limitations prevented training of neural networks with many layers, and progress in DL was slow. The emergence of faster and more powerful processors, along with the availability of large amounts of labeled data, led to a resurgence of interest in DL in the early 2000s. 
\subsubsection{Convolutional neural networks (CNNs)}
Researchers began to develop more sophisticated NN architectures, such as convolutional neural networks (CNNs) and recurrent neural networks (RNNs), that could learn from complex and high-dimensional data, such as images, leading to image recognition using a so-called deep convolutional neural network (DCNN)~\cite{hinton2015deep}. The employed convolutional layers are particularly well-suited for image data, as they use a set of learnable filters to convolve over the image, detecting various features such as edges, corners, or textures. Since then, DL has become one of the most active areas of research in artificial intelligence (AI), with applications in areas such as computer vision, natural language processing, speech recognition, and robotics.
\newline
DL has been used for many machine learning tasks of images, including classification, regression, and segmentation. The most popular DL architecture for image segmentation is the U-Net which is a fully convolutional neural network that was introduced in 2015 by Ronneberger et al~\cite{ronneberger2015_Unet}. The U-Net consists of an encoder network and a decoder network. The encoder network consists of several convolutional and pooling layers that decrease the spatial dimension of the input images while simultaneously increasing the number of feature maps. The decoder network is made up of convolutional and upsampling layers that restore the spatial dimensions of the resulting segmentation map and simultaneously decrease the number of feature maps. The U-Net utilizes skip connections to combine low-level features from the contracting path with high-level features from expanding path for preserving spatial resolution in the output. 
\newline
CNNs are one of the two most often used types of predicitve models, besides generative models of the GAN family (see below). In particular, the UNet is the most popular CNN architecture used for digital staining~\cite{44,46,54,57,60,62,69,88,101,103,104,110,120,133}.

\subsubsection{Generative models}
Generative adversarial networks (GANs) have revolutionized the field of DL by enabling the generation of realistic data samples. The first GAN was proposed by Ian Goodfellow in 2014~\cite{goodfellow2014_GANs}, and consisted of one generator and one discriminator. The generator produces fake data samples, while the discriminator tries to distinguish between real and fake data samples. The training process involves the two networks playing a min-max game, with the generator trying to fool the discriminator into classifying its fake samples as real, while the discriminator tries to correctly classify the samples. While GANs work most of the time, there is no guarantee that the generator will produce images that actually look like the input dataset. To address this issue, researchers have proposed various modifications to the GAN architecture, such as the conditional GAN, which adds class labels to the generator~\cite{goodfellow2014_GANs}, and the CycleGAN, which consists of two generators and one discriminator~\cite{liu2017_unsupervised_map2satelite}. Within this GAN family, the conditional GAN architecture of 'Pix2Pix' is the most commonly used for digital staining~\cite{57,81,91,93,94,105,115,118,119,129,130}. 
\newline
The CycleGAN has gained popularity in recent years due to its ability to translate between different modalities without the need for paired datasets with labels. Instead, the CycleGAN uses a cycle consistency loss to ensure that the generated output is consistent with the input data~\cite{liu2017_unsupervised_map2satelite}. This has enabled researchers to apply the CycleGAN to a wide range of tasks, such as predicting H\&E stain from photoacoustic microscopy images~\cite{128} and improving periodic-acid-Schiff-stained renal tissue for whole slide image segmentation~\cite{109}. In addition, translations between different stains have been proposed, like , transferring between Papanicolaou and Giemsa stains~\cite{92}. Moreover, CycleGAN approaches have also been used to improve periodic-acid-Schiff-stained renal tissue for whole slide image segmentation~\cite{109}, as well as to predict color brightfield images and antibody conjugates stained mouse kidney slides from monochromatic phase images reconstructed with Fourier ptychography~\cite{41}. In general, CycleGAN approaches have proven to be versatile and flexible, allowing for the translation between various modalities, making it easier to acquire the data required for medical diagnostics and research. One recent development in the field is the introduction of saliency maps, which have been used to improve the performance of unsupervised models for image transformation tasks. For example, an unsupervised model named Unsupervised content-preserving Transformation for Optical Microscopy (UTOM) uses a saliency constraint to learn the mapping between different histology stains~\cite{73}. 
\newline
One of the major challenges with GANs is the problem of "hallucination"~\cite{cohen2018hallucinate}. Hallucination occurs when the generator produces synthetic data that do not correspond to the input data distribution. In other words, the discriminator is still fooled by synthetic data that either shows realistic looking artifacts (e.g., a digitally stained cell, when there is no actual cell in that region) or by synthetic data that deletes features (e.g., cells) that are actually present in the real data. This problem can arise when the training data are limited or when the input data are highly variable. Hallucination can be problematic, particularly in the medical domain. It is difficult to detect when a GAN is hallucinating, as the synthetic data may look plausible to the human eye. To mitigate the problem, researchers have proposed various techniques, such as incorporating regularization terms in the GAN loss function~\cite{salimans2016improved}, using pre-training of the generator~\cite{goodfellow2014_GANs}, or using multiple discriminators~\cite{durugkar2016generative}. However, the problem of hallucination remains a challenging issue in GAN training, particularly when working with limited or highly variable data, as discussed below.

\subsection{Loss functions}
\label{sec:losses}
For deep learning-based virtual staining, loss function selection is one of the most important aspects of the neural network designing. 
Similar to other DL applications, the most commonly used loss functions are mean absolute error (MAE) or L1 loss, the mean squared error (MSE), which takes the L2-norm penalty, and cross-entropy. The innovations in the fields of convolutional networks, the U-Net and generative models, were accompanied by a series of new quantitative metrics for image similarity, such as Wasserstein loss~\cite{frogner2015learning}, structural similarity index (SSIM)~\cite{brunet2011mathematical} or multi-scale SSIM (MS-SSIM)~\cite{wang2003multiscale}. In addition to MSE and MAE, these metrics are used frequently for training and/or performance evaluation in digital staining. However, employing a single loss function may lead to performance degradation. MAE keeps brightness and color unchanged, but assumes that the influence of noise and the local characteristics of the image are independent~\cite{91}, MSE tends to generate blurry results~\cite{22}. SSIM became popular since it tends to produce results that are closer to the human visual system in terms of brightness, contrast, structure, and resolution~\cite{56}. Additionally, SSIM can detect high-level structural errors~\cite{73}. However, we note that the MS-SSIM loss can lead to brightness changes and color deviations~\cite{91}. Therefore, there is a range of customized metrics designed for specific tasks, as well as the combination of multiple basic loss functions~\cite{15,59,68,73}. In Table~\ref{tab:table_losses}, we summarize formal definitions and featured references for all loss functions. 

\begin{table}[]
    \centering
    \caption{Selected loss functions used for digital staining (DS) with featured references, not including adversarial losses. $O(x,y)$ represents the output image, $\hat{O}(x,y)$ represents the target image, $\mu$ represents the average value, $\sigma$ represents the standard deviation, $c_1$ and $c_2$ are stabilization constants used to prevent division by weak denominator, respectively.}
    \begin{tabular}{ p{3cm}|p{5.7cm}|p{3cm} }
       Metric & Formal definition & DS References \\
       \hline
       Mean Absolute Error (MAE) & $L_{MAE} = \sum_{x=1,y=1}^{X,Y}|O(x,y)-\hat{O}(x,y)|$ & \cite{28,69,91,106} \\
       & & \\
       Mean Squared Error (MSE) & $L_{MSE} = \sum_{x=1,y=1}^{X,Y}(O(x,y)-\hat{O}(x,y))^2$ & \cite{29,33,58} \\
      & & \\
       Cross-entropy & $L_{CE} = \sum_{x=1,y=1}^{X,Y}\hat{O}(x,y)log(O(x,y))$ & \cite{44,52,60,72}\\
       & & \\
       Structural Similarity Index Measure (SSIM) & $L_{SSIM} = \frac{(2\mu_o\mu_{\hat{o}}+c_1)(2\sigma_o\sigma_{\hat{o}}+c_2)}{(\mu_o^2 + \mu_{\hat{o}}^2+c_1)(\sigma_o^2 + \sigma_{\hat{o}}^2+c_2)}$ & \cite{29,33,58}\\
    \end{tabular}
    \label{tab:table_losses}
\end{table}

The invention of GANs and their wide use for digital staining, as discussed above, require more complex adversarial loss metrics that are composed of generator loss and discriminators loss. CycleGAN models~\cite{zhu2017unpairedcylcleGAN} typically contain two terms: the adversarial loss, to quantify the style match between target and generated images, and a cycle consistency loss $L_{cyc}(G,F)$, which prevents the learned mappings $G$ and $F$ from contradicting each other. Additional losses are also often incorporated into these basic terms, i.e., for regularization purposes. 

\section{Generation of paired images}
\label{section:generate_pairs}
Digital staining relies on paired images. Although some GAN-based techniques use unpaired data sets for unsupervised training, the majority of articles reviewed here still relies on paired images for supervised learning. Moreover, due to the "hallucination-gap" mentioned above, we postulate that paired images are a necessity at least for a trustworthy validation and performance evaluation of a given digital staining model.

\begin{figure}[ht!]
\begin{center}
\includegraphics[width=\linewidth]{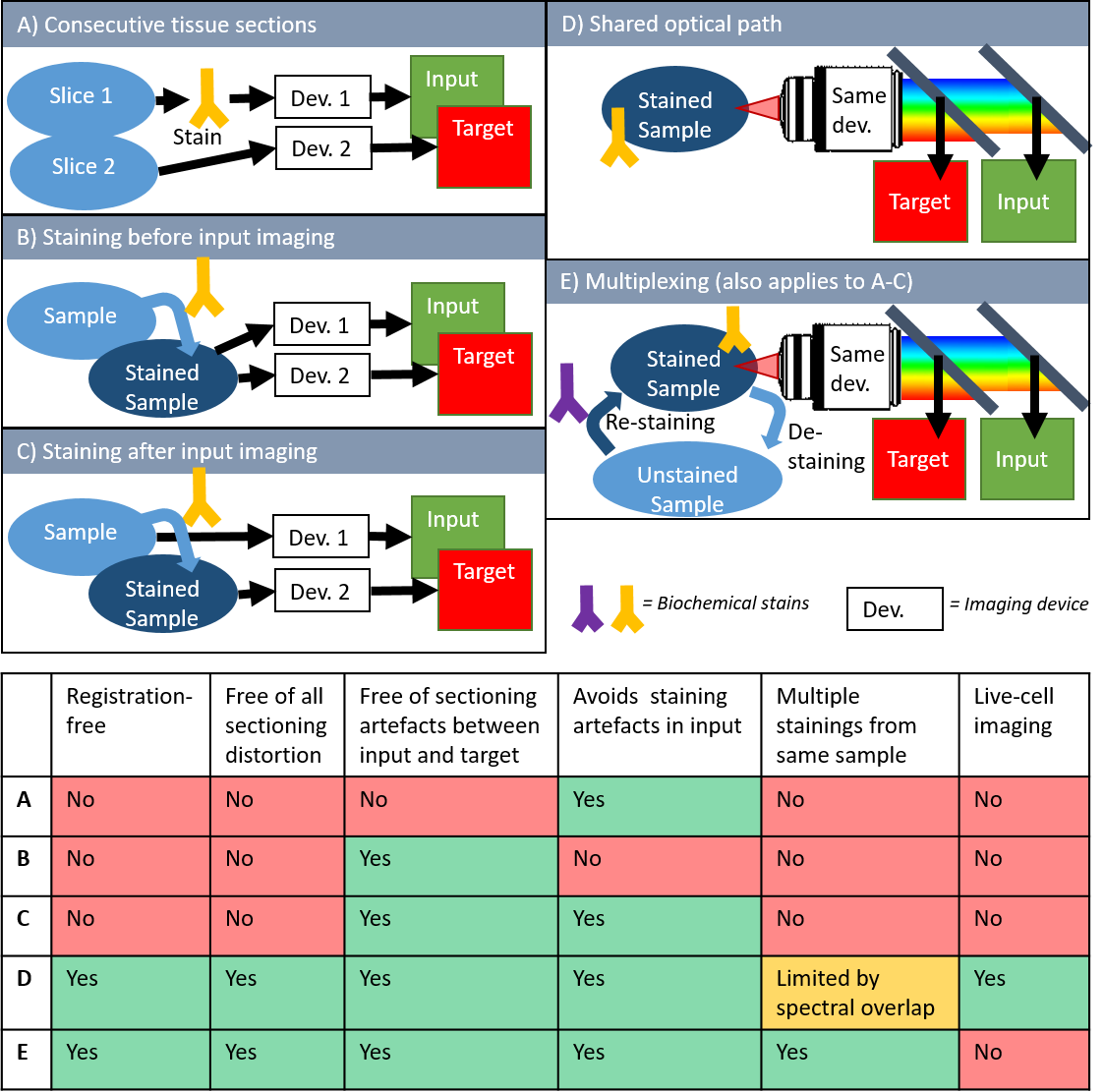}
\end{center}
\caption{\textbf{Generation of image pairs for the training of digital staining models} (A-E) Schematic workflow of the five different procedures. The table shows positive features (green), neutral features (orange) and negative features (red). }
\label{fig:fig4}
\end{figure}

The generation of these paired input and target images is as an important consideration in the practical implementation of digital staining. With the exception of earlier studies that used mathematical equations for color / style transfer~\cite{7, 10, 9, 11, 84, 14, 85, 23, 24, 75, 65} and most recent techniques that use semi-supervised or un-supervised ML models~\cite{41,50,70,73,124,128}, most approaches use paired images of input and target space for training. At the very least, a truthful validation of the output of trained digital staining models still requires paired images, even for conventional linear color translation or for modern unsupervised learning. Therefore, the process of sample preparation, staining protocol and sequence of imaging is also important for digital staining. Here, we have identified five main procedures, as also displayed in Figure~\ref{fig:fig4}:

\begin{itemize}
    \item cutting \textbf{consecutive tissue sections} from a block of FFPE tissue, and imaging each at a different device (e.g., one for label-free input and one for actual staining as target image). 
    \item the sample is \textbf{first stained and then imaged} consecutively by two different techniques (e.g., one for input and one for target imaging)
    \item the unstained sample is \textbf{first imaged} for the (label-free) input domain and is \textbf{then stained} for the target image domain 
    \item the choice of input contrast and target contrast allows for spectral separation between input and target images in the same \textbf{shared optical path}. 
    \item multiplex-staining or \textbf{de- and re-staining}. Here, the same sample is imaged multiple times with multiple different staining techniques. A previous set of stains is chemically removed or bleached, before the next set is applied.
\end{itemize}

The unique advantages and disadvantages of these techniques are summarized in the table in Figure~\ref{fig:fig4}. Please note that these limitations are only relevant for the generation of image pairs, and, therefore for the development/training and the verification of single digital staining models. While early studies relied mostly on working with consecutive sections (Fig.~\ref{fig:fig4}~A), imaging of the same tissue section is actually the preferred method of choice to remove sectioning artifacts between input and target. Ideally, staining of the target contrast is performed \textit{after} input imaging, although a few niche applications used a workflow where the sample was first stained (Fig-~\ref{fig:fig4}~B). Whenever different imaging platforms are used sequentially (Fig.~\ref{fig:fig4}~A-C), image registration is still essential (see section~\ref{sec_computational_models_preproc}). In contrast to that, techniques with shared optical path can omit the need for image registration, while also enabling digital staining of cell cultures without the presence of tissue sectioning artifacts for continuous digital staining of processes, like cell growths and cell-to-cell interaction (Fig.~\ref{fig:fig4}~D). Multiplexing of the staining protocol by using de-staining and re-staining protocols (Fig.~\ref{fig:fig4}~E), can maximize the amount of staining from a given sample (tissue section or cell culture). The main advantages and disadvantages of each technique are summarized in Fig.~\ref{fig:fig4}.

\section{Applications}
In this section, we categorized digital staining publications according to the type of sample and according to the field of application. As already mentioned in section~\ref{sec_basic_principle}, there are currently two main modes of operation. On the one hand, the use of fixed tissue sections is most common, either relying on FFPE sections~\cite{2, 1, 3, 5, 6, 7, 10, 8, 11, 107, 14, 66, 85, 18, 95, 108, 19, 20, 125, 28, 82, 22, 27, 38, 110, 29, 39, 70, 24, 36, 122, 71, 120, 35, 69, 121, 41, 80, 37, 73, 49, 53, 127, 133, 47, 65, 106, 45, 55, 57, 62, 93, 117, 126, 115, 116, 94, 112, 119, 100, 113, 102, 111, 129, 124, 131, 132} or on frozen tissue sections~\cite{9, 15, 50, 51, 128}. The second main field is digital staining for cell cultures~\cite{60, 58, 23, 72, 75, 68, 104, 33, 61, 52, 92, 99, 46, 101, 42, 59, 56, 91, 54, 88, 74, 114, 103, 118}, either from fixed or \textit{in vitro} cell samples~\cite{58,74,103,104}. Additional, but minor fields of application include the use of fresh, un-preserved tissue samples~\cite{48, 81} or even some \textit{in vivo} studies, e.g., on the skin~\cite{105}, in opthomological imaging of the retina~\cite{89} or during endoscopic imaging~\cite{83}.
\newline
The immediate goal of digital histopathology staining of tissue sections (sometimes termed 'pseudo-H\&E' staining) is to facilitate a wider use of label-free optical technologies by physicians and biomedical researchers, as it allows analysis of label-free images by a pathologist in the well-known and accepted histology image domain~\cite{Rivenson2020_Review,66}. Furthermore, it could allow the use of routine image analysis protocols that have been developed for conventional stainings, e.g., for surgical margin analysis~\cite{116} or white blood cell identification in blood smears~\cite{130}.
\newline
Compared to virtual histology staining of tissue sections and blood smears, digital staining of cells cultures offers entire new research applications that could otherwise not be investigated. IHC staining are unfeasible, especially if cells need to be kept alive. Even fluorescence antibodies stains can interfere with biological processes, if their molecular size is large. A common applications for digital staining of cell cultures is the distinction between live and dead cells using label-free imaging and digital staining~\cite{54,60,61,72,73}. The technique is also frequently applied to neurons~\cite{46,60,69,73}, where functional information from living cells is especially interesting and where actual staining can be particular challenging. The combination of label-free imaging and digital staining allowed the simultaneous use of an AI-based nucleus finding algorithm and an additional tracking algorithm, which was not possible to the traditional method, as fluorescent tracking can affect cell behavior~\cite{103}. This concept of combining digital staining with object detection, i.e., for nucleii detection is also used in other applications~\cite{132}. As already mentioned, digital staining of phase microscopy images enabled investigation of cell growth and cell division, where the translation model was trained on samples concurrently stained for the G1 and the S stage of the cell cycle~\cite{74}. The overlap of both signals could then indicates the G2 or M stage~\cite{74}. The concept was even extended to infer not only the staining procedure, but also 3D optical sectioning capability of confocal fluorescence microscopy based on non-confocal 3D quantitative phase images~\cite{133}. The approach was generalized for different fluorescence channels, different cell types and different magnifications~\cite{133}.

\section{Trends \& methods of good practice} 
\label{Diss_goodpractice}

\begin{figure}[ht!]
\begin{center}
\includegraphics[width=0.8\linewidth]{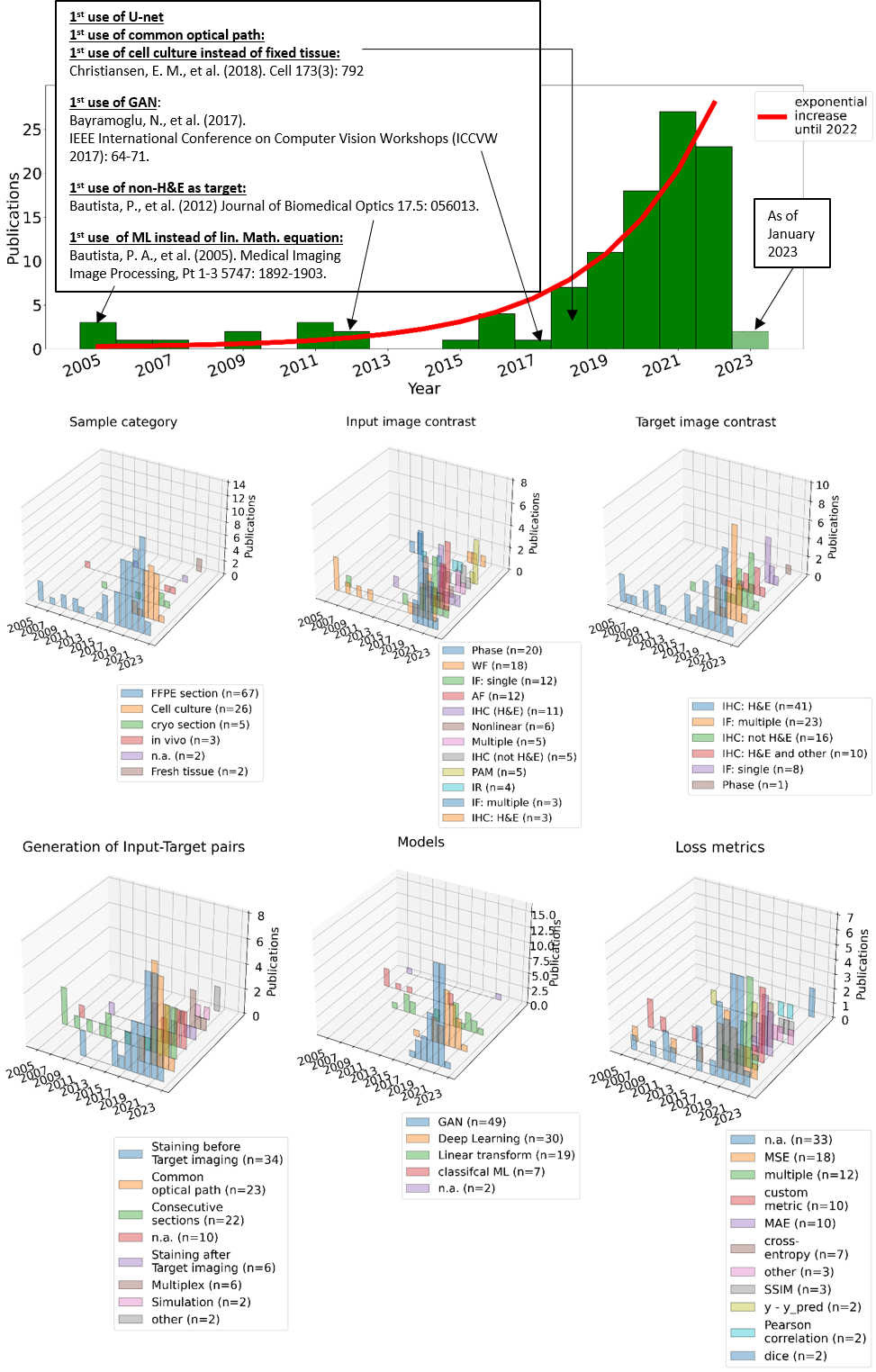}
\end{center}
\caption{\textbf{Historical trends in the field of digital staining}. (a) The total number of publications in the field (b-h) and grouped according to the categories. }
\label{fig:sup_fig3}
\end{figure}

Pillar and Ozcan identified several key advantages of virtual IHC staining over actual staining~\cite{pillar2022_shortReview_AF}, like a reduced time to perform staining, minimal manual labor, minimal stain variability, less hazardous waste composition of tissue fixatives, preservatives and staining dyes, no tissue disruption of the actual sample, no restrains to use multiple stains on a single slide, the chance to perform stain-to-stain transformation and a reduced chance for technical failures~\cite{pillar2022_shortReview_AF}. Although this review was only targeting the sub-genre of virtual histology staining, most of these advantages also generally apply to digital staining, as it is discussed here. An important addition to the field is the use of digital staining for cell cultures (both fixed and alive cells), as discussed above. In this case, digital staining offers additional advantages, such as an identification of functional stages (e.g., growth phase) without the biochemical binding of actual antibody stains, that could otherwise interfere with biological homeostasis and impact motion, growths or other aspects of relevance.
\newline
As digital staining was refined over the years, we can identify certain trends in this field (see supplementary figure~\ref{fig:sup_fig3}). While the first techniques mostly used linear color translation for pseudo-H\&E staining, computational tools became more powerful and the applications became more diverse over time. Nowadays, DL models, like the U-Net (since 2015), or GAN models (since 2016), are the most common models used for digital staining. At the same time, the range of applications has significantly expanded, coming from histological tissue sections to cell cultures (since 2018), multiplexed cell imaging (since 2018), or even advanced examples mentioned above, like live cell growth imaging~\cite{74} or inference of 3D confocal fluorescence from non-confocal phase images~\cite{134}. Similarly, the applied input imaging technologies diversified over time. Label-free modalities, like phase contrast (20/107 articles), wide-field (17/107 articles) and single-photon autofluorescence (12/107 articles) microscopy are still the most frequently used, making good use of digital staining to add computational specificity to these label-free technologies. However, digital staining is also used for stain-to-stain translations, e.g., with artificial stains (H\&E with 15/107 or IF with 12/107 articles) as input.
\newline
Based on these ongoing trends and the current state-of-the-art, we suggest the following methods of good scientific practice, when developing digital staining. Since each of these topics is an entire field of research in itself, we will only shortly address their relevance to the field of digital staining.

\begin{itemize} 
    \item \textbf{General feasibility:} As with most ML problems, one should consider first, whether the information content in the data, i.e., the input domain, is believed to be sufficient for the given task. More specifically, a good first question is if the general information in the input images is correlated with the one in the target domain. For instance, it might seem unfeasible to digitally stain cell nuclei (target) from images that only contain fluorescence of a membrane marker as input, if no additional information was used. On the other hand, it would seem quite feasible to perform DS of nuclei and membrane markers based on phase contrast images, as the contrast in optical phase is high for both nuclei and membranes. If a paired data set is already available, we suggest to test the general feasibility first by developing a model for simpler tasks, such as patch classification, object detection, or semantic segmentation.
     \item \textbf{Report uncertainties:} One of the main short-comings of the current state-of-the-art for digital staining is that fundamental uncertainties in input and in target data are usually not reported. As presented in this review, however, DS is a holistic approach that involves the entire pipeline of biology, imaging and ML. Simply reporting a performance metric of the ML task, is therefore insufficient, as those metrics assume a perfect ground-truth. However, target data from actual biochemical staining are always affected by the specificity of the molecular marker as fundamental uncertainty in the ‘ground truth’ (see section~\ref{Targets_specificity}). Similarly, imaging of inputs and targets is subject to the specific contrast mechanism, resolution and SNR of the respective imaging technology. Thus, we propose that \textbf{digital staining should always be embedded in the context of input and target uncertainties} of the actual stain as well as SNR of the imaging process to allow a fair evaluation of its performance. 
     \item \textbf{Generalizability:} there is a generalization gap~\cite{wagner2022more_reproducible} in DL and digital pathology, which also applies to digital staining. DS can often be very hardware-specific and can be prone to over-fitting. Therefore, it is essential to discuss generalizability. Ideally, one should take a hardware-agnostic approach when testing a DS pipeline. It is recommended to validate and test a system across different imaging systems and/or different tissue types to evaluate if it generalizes well. This can further be extended to evaluate the generalizability across different experimenters, different staining methods or different data sets. See references~\cite{58,60,106,126,118,135} for good examples. 
     \item \textbf{Choice of the right loss function:} After the selection of input and target technologies (which might be predefined for a given problem), the choice of the loss function is important. Different loss functions can emphasize different aspects of the image-to-image regression task, e.g., high-level structural errors (SSIM), absolute errors at the pixel level (peak signal-to-noise ratios - PSNR), brightness and color (MAE) or custom loss functions (see section~\ref{sec:losses} for more details). 
    \item \textbf{Image inspection and decision visualization:} Besides the mere reporting of loss curves and performance metrics, it is indispensable to visually inspect and report the actual target and prediction images. Although the above mentioned loss functions are suited for training and quantitative performance comparison, some can be ill-suited to detect hallucinations~\cite{cohen2018hallucinate}, artifacts or other localized prediction errors in the images. Moreover, decision visualization, like occlusion maps, Shapley values or perturbation studies can inform the researcher about features that are particularly important to the learning process. This can not only support de-bugging during the development of DS, but it can also offer valuable scientific feedback e.g., to understand which parts of an input image are particularly relevant to predict a certain target.
    \item \textbf{Interpretability:} similar to the point above, ML models can often lack interpretability, which prevents identification of biases and can thereby reduce generalizability. Interpretability is especially relevant for digital staining to prevent false halluzinations from overfitting models. A good rule of thumb is that simpler models with a small number of parameters are more interpretable. Furthermore, it is preferred to create more interpretable models from the beginning instead of post-hoc explanations of complicated models~\cite{Rudin2019}. 
    \item \textbf{Availability of code \& data:} Whenever possible, it is recommended to make code and data available to other researchers, according to the FAIR principle (Findability, Accessibility, Interoperability, and Reuse of digital assets). This enhances trustworthiness and transparency of the general scientific procedure and further enables other researchers to test new approaches, especially since good data sets of paired images might be a bottleneck for many ML researchers. Positive examples, where code \textit{and} data were made public are~\cite{15,47,56,58,59,60,68,73,106,118}
    \item \textbf{Multi-modal imaging:} Combinations of different contrast mechanisms for a richer information content are more robust. Examples include FPM as a natural combination of amplitude and phase~\cite{41}, dark field reflectance and autofluorescence (DRUM)~\cite{131} or complementary nonlinear techniques, like CARS, SHG  and two-photon AF~\cite{18,50,66}. 
\end{itemize}

\section{Caution \& pitfalls}
\label{Diss_pitfalls}
As antithesis to the methods of good practice discussed above, we identify certain pitfalls that can reduce the overall success of digital staining (i.e., prediction performance, robustness, validity, computation time, and required number of examples).
Generally, we consider the error analysis of digital staining to be not fully developed yet. While most articles in the field do a very good job to report a growing collection of ML performance metrics, a holistic error analysis of the entire process, is not part of the state-of-the-art. We postulate that error analysis for digital staining should include modeling uncertainties (ML performance metrics, training curves but also errors from pre-processing, e.g., image registration) \textit{and} biological uncertainties (binding specificity, purity of cell cultures, contamination, bleaching of fluorescence), as well as optical uncertainties (contrast, resolution, SNR). Moreover, high uncertainties in input contrast and target ‘ground truth’ will remain undetected, if data from the same general population (e.g., the same target stain and same imaging system) are used for the validation of predictions and for the overall performance evaluation.

\section{Outlook and anticipations}
\label{Diss_trends}
In this final chapter, we take a few educated guesses on potential future trends for this field, based on our in-depth analysis of the historical development and current state-of-the-art of digital staining.

Generally, medical diagnostics in remote and resource-limited settings would greatly profit from a low-cost, stainless approach like digital staining. When applied to simple and robust systems, like portable white-light or phase contrast microscopes, this could enable reasonable diagnostic yield from inexpensive hardware. Although label-free technologies, like MPM, CARS, PAM, FPM and others, are growing fields of research, digital staining is currently still under-investigated for these emerging techniques. Thus, we foresee a further implementations of digital staining for these more advanced optical contrast mechanisms. Furthermore, we believe that the input and target images with rely more on multiple different stains and/or mpIF, which was shown to have a higher accuracy in diagnostic prediction as compared to single stainings~\cite{lu2019comparison}.
\newline
In the branch of ML models that are used for digital staining, several innovations can be imagined. 
For once, multi-task learning is an emerging concept that is being used for ML in optical microscopy. As it was already realized for digital staining with auxiliary tasks~\cite{52}, it will likely become more relevant for this field in the future. The concept of multi-domain image translation, i.e., training a single model to learn mappings among multiple domains, was already implemented in a larger number of publications~\cite{58,71,91,112,113}.
In a similar fashion, physics-informed learning and integration of  prior information or simulation data into the learning process are interesting concepts in modern ML research. Since these are well-suited to increase robustness and generalizability, they would probably be able to address several challenges in modern digital staining, as in the case of hallucinations. This concept has not really found its way to digital staining yet, exceptfor publications that employed simulations to improve the training process~\cite{135} or that modeled the microscope’s point spread function in the learning of an adversarial neural network to improve digital staining~\cite{118}. The trend in the development of new ML models, from classical ML over DL to GAN models, is likely to continue and to produce entirely new concepts for ML models. One potential candidate is Adversarial Diffusion Models. These are already used to translate between MRI and CT data~\cite{ozbey2022_Adversarial_Diffusion_Models}, which is a very similar problem to digital staining in optical microscopy.
\newline
The continuation of current trends as well as potential innovations in the field will very likely result in a series of exciting new applications for digital staining. Although digital staining of histology sections has shown to facilitate easier, faster and potentially more accurate clinical diagnosis in several research publications, a full FDA approval as medical product will remain challenging, due to extensive documentation requirements and current technical limitations. We believe that this technique is currently more interesting for cell cultures, as discussed in this review. Since this use-case does not imply sensitive patient data or critical decisions on clinical diagnosis, a commercialization in the biotechnology sector is probably more feasible. The technique of 3D fluorescent labeling based on phase microscopy was already patented~\cite{101}. Following this trend, digital staining could potentially be used for organoids, that gained a lot of popularity in the recent years. 
\newline
In the long-term future, however, clinical applications of digital staining would not only be limited to tissue sections but could become a vital tool for clinical \textit{in vivo} imaging. Currently, DL is already used to improve image quality in endomicroscopes~\cite{guan2022_DL_Endomicroscope}, and endoscopic or endomicroscopic implementations are already available for many imaging technologies and optical contrast mechanisms mentioned in this review. This next step of digital staining, however, needs to be accompanied by designing more robust, generalizable and interpretable models, as discussed above. This point was also identified by Jiang et al., who mentioned the problems of variable clinical factors regarding imaging microscopes, staining techniques, patch extraction, and selection and stated that "To address this issue, designing more robust architectures can make the model less dependent on data quality in digital medicine."~\cite{Computational_Cytology_review}. 
\newpage

\beginsupplement
\section{Supplementary Materials \& Analysis of literature}

 \begin{figure}
\begin{center}
\includegraphics[width=\linewidth]{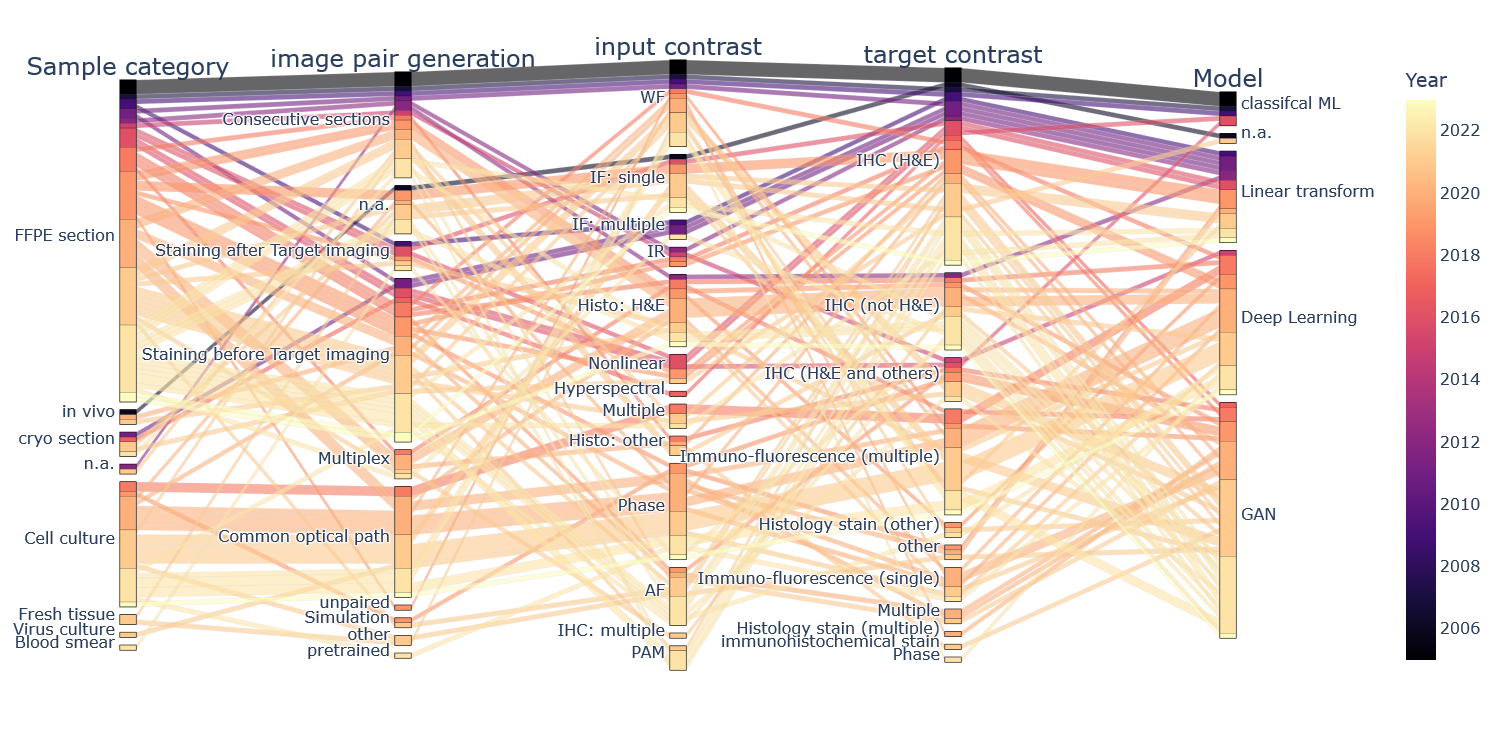}
\end{center}
\caption{\textbf{Parallel categories with connections} All reviewed articles as parallel and linked categories. The year of each publication is color-coded. An interactive version of this plot is available as supplementary html file.}
\label{fig:sup_fig1}
\end{figure}

\begin{figure}[ht!]
\begin{center}
\includegraphics[width=\linewidth]{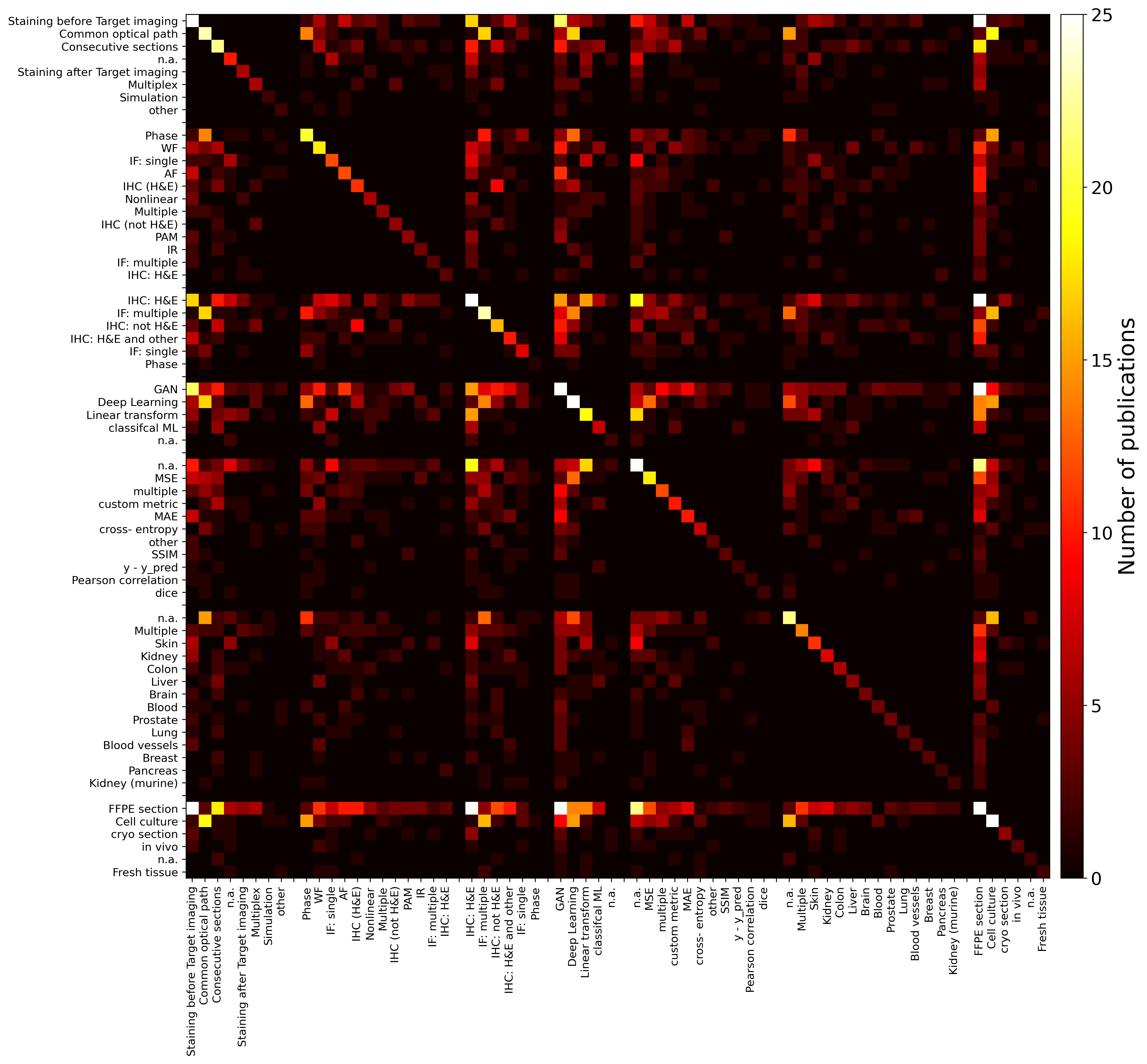}
\end{center}
\caption{\textbf{All combinations of different categories} as heatmap of  with the color-coded number of publications. Extended version of Figure~\ref{fig:fig2}}
\label{fig:sup_fig2}
\end{figure}

\section*{Methods for selecting literature}
\section*{Literature review}
For the literature data base in this review, 107 articles between 2005 and January 2023 were were reviewed and categorized. We considered peer-reviewed articles of above two pages length, not including short conference abstracts or un-reviewed preprints. Articles were considered as digital staining, if an image-to-image regression for microscopy images in different contrast domains was carried out. 
Articles that performed conventional segmentation tasks or stain normalization (i.e.,  transfer from one domain to the same domain) are not considered. The date of acceptance was used as time stamp. If that was not available, the date of publication was used. The full data base of the all reviewed articles that were used for the figures in this paper is available as supplementary material. The keyword search on google scholar contained the followed keywords and possible permutations thereof: virtual fluorescence, virtual staining, in silica label, computational specificity, computational stain, digital stain, in silico stain, pseudo H\&E, in silico label.

\section*{Visualizations}
Literature data in Figure~\ref{fig:fig2} and in supplementary Figures  \ref{fig:sup_fig2} and \ref{fig:sup_fig3} were handled using the pandas library and plotted in python. Supplementary Figure \ref{fig:sup_fig1} was generated using $plotly.express.parallel\_categories$. An interactive version of this plot is available as supplementary file.

\section*{Acknowledgments}
This project was supported by the European Union’s Horizon 2022 Marie Skłodowska-Curie Action no. 101103200, project MICS. 
\section*{Data availability}
The full data base of all reviewed articles and their respective categorizations is available as supplementary material. 

\section*{Author contributions statement}
LK conceptualized the work, performed literature review, categorized all reviewed publications, generated figures and wrote the manuscript with input from all co-authors. XL, KL, LB and OF provided input on biochemical staining as target images and on multiplexed digital staining. AM, SX, AC and KK provided input on DL and GAN models. SJ and GZ provided input on loss functions. SX, KCL and SAL provided input on phase contrast imaging. KZ provided input on white-field imaging. SJ, KL, MA and RH provided input on methods of good practice, pitfalls and potential trends.

\section*{Conflict of interest}
The authors declare no conflict of interest.

\bibliographystyle{unsrt}  
\bibliography{References}

\end{document}